\documentclass{ws-ijmpd}

\begin{document}

\markboth{CHANG-BO SUN,JIA-LING WANG and XIN-ZHOU LI} {Instructions
for Typing Manuscripts (Viscous Cardassian universe)}

\catchline{}{}{}{}{}
%
%%%%%%%%%%%%%%%%%%%%%%%%%%%%%%%%%%%%%%%%%%%%%%%%%%%%%%%%%%%%%%%%%%%%

\title{Viscous Cardassian universe}

\author{CHANG-BO SUN}
\address{Shanghai United Center for Astrophysics(SUCA),
 Shanghai Normal University, 100 Guilin Road, Shanghai 200234,China}
\author{JIA-LING WANG}
\address{Department of Physics, East China University of Science and Technology, Shanghai 200237, China}
\author{XIN-ZHOU LI}
\address{Shanghai United Center for Astrophysics(SUCA),
 Shanghai Normal University, 100 Guilin Road, Shanghai 200234,China\\
 kychz@shnu.edu.cn}
\maketitle

\begin{history}
\received{Day Month Year}
\revised{Day Month Year}
\comby{Managing Editor}
\end{history}

\begin{abstract}
The viscous Cardassian cosmology is discussed, assuming that there
is a bulk viscosity in the cosmic fluid. The dynamical analysis
indicates that there exists a singular curve in the phase diagram of
viscous Cardassian model. In the viscous PL model, the
equation-of-state parameter $w_{k}$ is no longer a constant and it
can cross the cosmological constant divide $w_{\Lambda}=-1$, in
contrast with same problem of the ordinary PL model. Other models
possess with similar characteristics. For MP and exp models, $w_{k}$
evolves more near $-1$ than the case without viscosity. The bulk
viscosity also effect the virialization process of a collapse system
in the universe: $\frac{R_{vir}}{R_{ta}}$ is increasingly large when
the bulk viscosity is increasing. In other words, the bulk viscosity
retards the progress of collapse system. In addition, we fit the
viscous Cardassian models to current type Ia supernovae data and
give the best fit value of the model parameters including the bulk
viscosity coefficient $\tau$.
\end{abstract}

\keywords{Bulk viscosity; Cardassian universe;cosmological dynamics;
virialization.}

\section{Introduction}
 The current accelerating expansion of the universe indicated by
  the astronomical measurements from high redshift supernovae\cite{supernovae} as
  well as accordance with other observations such as the
  Boomerang/Maxima/WMAP data\cite{Boomerang/Maxima/WMAP} and galaxy power spectra\cite{galaxy power spectra} becomes one of
  the biggest puzzles in the research field of cosmology. One
  popular theoretical explanation approach is to assume that there
  exists a mysterious energy component, dubbed dark energy, with
  negative pressure, or equation of state with $w=p/\rho <0$ that
  currently dominates the dynamics of the universe(see\cite{quintessence} and references therein).
   Such a component makes up $70\%$ of the energy density of
  the universe yet remains elusive in the context of general
  relativity and the standard model of particle physics. The most
  natural dark energy candidate is a cosmological constant which
  arises as the result of a combination of general relativity(GR)
  and quantum field theory. However, its theoretical value is
  between $60-120$ orders of magnitude higher than the observed
  value. An alternative candidate is a slowly evolving and spatially
  homogeneous scalar field, referred to as "quintessence" with
  $w>-1$\cite{quintessence} and "phantom" with $w<-1$\cite{phantom}, respectively. Since current
  observational constraint on the equation of state of dark energy
  lies in a relatively large around the so-called cosmological
  constant divide $w_{\Lambda}=-1$, it is still too early to rule out any
  of the above candidates. However, the expectation of explaining
  cosmological observations without requiring new energy component
  is undoubtedly worth of investigation. GR is very well examined in
  the solar system, in observation of the period of the binary
  pulsar, and in the early universe, via primordial nucleosynthesis,
  but no one has so far tested in the ultra-large length scales and
  low curvatures characteristic of the Hubble radius today.
  Therefore, it is a priori believable that Friedmann equation is
  modified in the very far infrared, in such a way that the universe
  begins to accelerate at late time. Freese and Lewis\cite{K.Freese2002} construct
  so-called Cardassian universe models that incarnates this hope.
  The Cardassian universe is a proposed modification to the
  Friedmann equation in which the universe is flat and accelerating,
  and yet contains only matter(baryonic or not) and radiation. But
  the ordinary Friedmann equation governing the expansion of the
  universe is modified to be
  \begin{eqnarray}\label{Modified FRW eq}
&&H^{2}\equiv \left(\frac{\dot{a}}{a}\right)^{2}=\frac{8\pi G}{3}
g(\rho)
\end{eqnarray}
where $\rho$ consists only of matter and radiation, $H$ is the
Hubble "parameter" which is a function of time, $a$ is the scale
factor of the universe, and $G=1/m_{pl}^{2}\equiv
\frac{\kappa^{2}}{8\pi}$ is the Newtonian gravitational constant.
Note that as required by inflation scenario and observations of CMB,
the geometry of the universe is flat, therefore, there are no
curvature terms in the above equation. The function $g(\rho)$
returns to ordinary Friedmann equation at early times, but that
takes a distinct form that drives an accelerated expansion in the
recent past of the universe at $z<\mathcal {O}(1)$. In Cardassian
models, we simply set the cosmological constant $\Lambda =0$, and
the only compositions are matter and radiation. A possible
interpretation of Cardassian arose from consideration of braneworld
scenarios, in which our observable universe is a 3-dimensional brane
embedded in a higher dimensional universe. An alternative
interpretation had been also discussed\cite{P.Gondolo}, in which one
developed a 4-dimensional fluid description of Cardassian cosmology.
The observational constraints of Cardassian models have been
extensively studied\cite{observational constraints of Cardassian}.
The simplest model is power law Cardassian modle(PL) with an
additional term $\rho^{n}$, which satisfies many observational
constrains: the first Doppler peak of CMB is slightly shifted, the
universe is rather older, and the early structure formation($z>1$)
is unaffected. Furthermore, one can consider other forms of
$g(\rho)$ including the modified polytropic model(MP) and the
exponential model(exp)
 \begin{eqnarray}\label{models}
&&g(\rho)=\left\{\begin{array}{ll}
\rho\left[1+\left(\frac{\rho}{\rho_{card}}\right)^{n-1}\right],&\textrm{for PL}\\
\rho\left[1+\left(\frac{\rho_{card}}{\rho}\right)^{qn}\right]^{\frac{1}{q}},
 & \textrm{for MP}\\
 \rho \exp\left[\left(\frac{\rho_{card}}{\rho}\right)^{n}\right],& \textrm{for
 exp}
 \end{array}\right.
\end{eqnarray}
where $\rho_{card}$ is a characteristic constant energy density and
$q$ and $n$ are two dimensionless positive constants\cite{Exp Car}.
Obviously, at early times, $\rho$ is much larger than the
characteristic energy density $\rho_{card}$, $g(\rho)\rightarrow
\rho$, i.e. Eq.(\ref{Modified FRW eq}) recovers the ordinary
Friedmann equation.

  On the other hand, the dissipative effects, including both bulk
and shear viscosity, are supposed to play a very important role in
the astrophysics (see Ref.\refcite{astro viscous} and references
therein) and the nuclear physics (see Ref.\refcite{nucl viscous} and
references therein). Under the conditions of spatial homogeneity and
isotropy, a scalar bulk viscous pressure is the solely admissible
dissipative phenomenon. The viscosity theory of relativistic fluids
was first suggested by Eckart\cite{Eckart} and Landau and
Lifshitz\cite{Landau}, who considered only first-order deviation
from equilibrium, which leads to parabolic differential equations
and hence to an infinite speed of propagation for heat flow and
viscosity, in contradiction with the principle of causality. The
relativistic second-order theory was founded by Israel\cite{Israel},
and has also been used in the evolution of the
universe\cite{T.Harko}. However, the character of the evolution
equation is very complicated in the framework of the full causal
theory. Therefore, the conventional theory\cite{Landau} is still
applied to phenomena which are quasi-stationary, i.e., slowly
varying on space and time scales characterized by the mean free path
and the mean collision time. In the case of isotropic and
homogeneous cosmologies, the dissipative process can be modelled as
a bulk viscosity $\zeta$ within a thermodynamical approach. Some
original works on the bulk viscous cosmology were done by Belinsky
and Khalatnikov\cite{Belinsky}. The bulk viscosity introduces
dissipation by only redefining the effective pressure, $p_{eff}$,
according to $p_{eff}=p-3\zeta H$ where $H$ is the Hubble parameter.
The condition $\zeta >0$ assures a positive entropy production in
conformity with the second law of thermodynamics\cite{Zimdahl}. We
are interested in two solvable cases: (i)
$\zeta=\sqrt{3}\kappa^{-1}\tau H$, where $\tau$ is a constant. This
assumption implicates that $\zeta$ is directly proportional to the
divergence of the cosmic fluid's velocity vector. Therefore, it is
physically natural, and has been considered previously in an
astrophysical context\cite{Gr0n}; (ii) $\zeta=\tau
\left(g(\rho)\right)^{\alpha+\frac{3}{2}}$. This dependence is more
complicated, but one can see in the following some interesting
results obtained. Obviously, case (i) is equivalent to one with
$\alpha =-1$ of case (ii).

  In this paper, we consider a viscous Cardassian cosmological model
  for the expanding universe, assuming that there is bulk viscosity
  in the cosmic fluid. The dynamical analysis indicates that there
exists a singular curve in the phase diagram for various models. The
numerical result show that $\frac{R_{vir}}{R_{ta}}$ is increasingly
large when the bulk viscosity is increasing. Furthermore, we fit the
viscous Cardassian models to current SNeIa data and give the best
fit values of the parameters including the bulk viscosity
coefficient $\tau$.

\section{The model}
 In the viscous Cardassian model, the modified Friedmann equation
  is described by Eq.(\ref{Modified FRW eq}). For the late-time
  evolution of the universe we neglect the contribution of
  radiation, so that  Eq.(\ref{Modified FRW eq}) is reduced to
 \begin{eqnarray}\label{Card FRW eq}
&&H^{2}=\frac{\kappa^{2}}{3} g(\rho_{m})
\end{eqnarray}
where $\rho_{m}$ is the energy density of matter, which keeps
conserved during the expansion of the universe, i.e.,
 \begin{eqnarray}\label{matter energy conservation eq}
&&\dot{\rho}_{m}+3H(\rho_{m}-3H\zeta_{m})=0
\end{eqnarray}
where $\zeta_{m}$ is the bulk viscosity for the matter energy
density $\rho_{m}$. In the $\zeta_{m}=0$ case, the evolution of
matter takes the ordinary manner $\rho_{m}=\rho_{m,0}(1+z)^{3}$
where $\rho_{m,0}$ is the present value of energy density of matter.
However, the evolution of matter is not likely power law of $(1+z)$
in the $\zeta_{m}\neq0$ case. Similarly, the conservation equation
of total energy density can be written as
 \begin{eqnarray}\label{total energy conservation eq}
&&\dot{g}(\rho_{m})+3H\left[g(\rho_{m})+p-3H\zeta\right]=0
\end{eqnarray}
where $\zeta$ is the bulk viscosity for total energy density
$g(\rho_{m})$. Combing Eqs.(\ref{matter energy conservation eq}) and
(\ref{total energy conservation eq}), it is easy to check that
$\zeta =\frac{\partial g(\rho_{m})}{\partial \rho_{m}}\zeta_{m}$.
Following Ref.\refcite{P.Gondolo}, we take the energy density to be
the sum of two terms:
 \begin{eqnarray}\label{two terms}
&&g(\rho_{m})=\rho_{m}+\rho_{k}
\end{eqnarray}
where $\rho_{k}$ is a Cardassian contribution. The thermodynamics of
an adiabatically expanding universe tell us that pressure of
Cardassian contribution is
\begin{eqnarray}\label{card pressure}
&&p_{k}=\rho_{m}\frac{\partial g(\rho_{m})}{\partial
\rho_{m}}-g(\rho_{m}),
\end{eqnarray}
and $p=p_{k}$ because of $p_{m}=0$. From Eqs.(\ref{two terms}) and
(\ref{card pressure}), we have the equation of state
\begin{eqnarray}\label{eq of state}
&&w_{k}=\frac{p_{k}-3H\zeta}{g(\rho_{m})-\rho_{m}}.
\end{eqnarray}
Furthermore, we have
\begin{eqnarray}\label{acceleration eq}
&&\dot{H}+H^{2}=\frac{\ddot{a}}{a}=-\frac{\kappa^{2}}{6}\left[g(\rho_{m})+3(p_{k}-3H\zeta)\right]
\end{eqnarray}
which is the acceleration equation of the cosmological expansion.
Using Eqs.(\ref{Card FRW eq}),(\ref{total energy conservation eq})
and (\ref{card pressure}), we obtain
\begin{eqnarray}\label{evolution eq}
&&a\frac{dg(\rho_{m})}{da}+3\rho_{m}\frac{dg(\rho_{m})}{d\rho_{m}}-3\kappa
\sqrt{3g(\rho_{m})}\zeta=0.
\end{eqnarray}
We shall be interested in the evolution of the late universe, from
$t=t_{0}$ onwards, where $t_{0}$ is the initial time and the
corresponding scale factor $a(t_{0})=a_{0}$ and energy density
$\rho_{m}(t_{0})=\rho_{m,0}$. From Eq.(\ref{evolution eq}), we have
\begin{eqnarray}\label{evolution of a}
&&a=a_{0}\left(\exp\left[\int_{\rho_{m,0}}^{\rho_{m}}\frac{dg(\rho_{m})}{3\kappa\sqrt{3g(\rho_{m})}\zeta-3\rho_{m}\frac{dg(\rho_{m})}{d\rho_{m}}}\right]\right)
\end{eqnarray}
which is the general relation between the cosmological scale factor
$a$ and the energy density $\rho_{m}$.

  In what follows, we focus on the viscous Cardassian model, as the
  cosmological dynamics are analytically solvable when we choose $\zeta=\sqrt{3}\kappa^{-1}\tau
H$ where $\tau$ is a constant. The scale factor is given by
\begin{eqnarray}\label{a expression}
&&\frac{a}{a_{0}}=\frac{f(\Omega_{m})}{f(\Omega_{m,0})}
\end{eqnarray}
where $\Omega_{m}$ is the density parameter of matter and
$\Omega_{m,0}$ is the present value of $\Omega_{m}$. For the PL
model, we have
\begin{eqnarray}\label{PL fOmega}
&&f(\Omega_{m})=\left[\frac{\Omega_{m}^{(\sqrt{3}\kappa\tau-1)n}(\Omega_{m}-1)^{n-\sqrt{3}\kappa\tau}}{[n(\Omega_{m}-1)-\Omega_{m}+\sqrt{3}\kappa\tau]^{\sqrt{3}\kappa\tau(n-1)}}\right]^{\frac{1}{3(n-1)(n-\sqrt{3}\kappa\tau)(\sqrt{3}\kappa\tau-1)}}
\end{eqnarray}
and the equation of state
\begin{eqnarray}\label{PL eq of state}
&&w_{k}=\frac{(1-n)(\Omega_{m}-1)-\sqrt{3}\kappa\tau}{1-\Omega_{m}}.
\end{eqnarray}
In the late time of universe, the new term of PL model is so large
that the ordinary first term $\rho_{m}$ can be neglected. In other
words, we have $\Omega_{m}\ll 1$, so that the expansion is
superluminal (accelerated) for $n<\frac{2}{3}+\sqrt{3}\kappa\tau$.

  For the MP model, we have
\begin{eqnarray}\label{MP fOmega}
&&f(\Omega_{m})=\left[\frac{\Omega_{m}^{q(\sqrt{3}\kappa\tau-1)(n-1)}(\Omega_{m}^{q}-1)^{n+\sqrt{3}\kappa\tau-1}}{(\sqrt{3}\kappa\tau+n-1-n\Omega_{m}^{q})^{\sqrt{3}\kappa\tau
n}}\right]^{\frac{1}{3qn(1-\sqrt{3}\kappa\tau)(\sqrt{3}\kappa\tau+n-1)}}
\end{eqnarray}
and
\begin{eqnarray}\label{MP eq of state}
&&w_{k}=\frac{n(\Omega_{m}^{q}-1)-\sqrt{3}\kappa\tau}{1-\Omega_{m}}.
\end{eqnarray}

\begin{figure}[!htbp]
\centering
 \includegraphics[height=0.9in,width=1.4in]{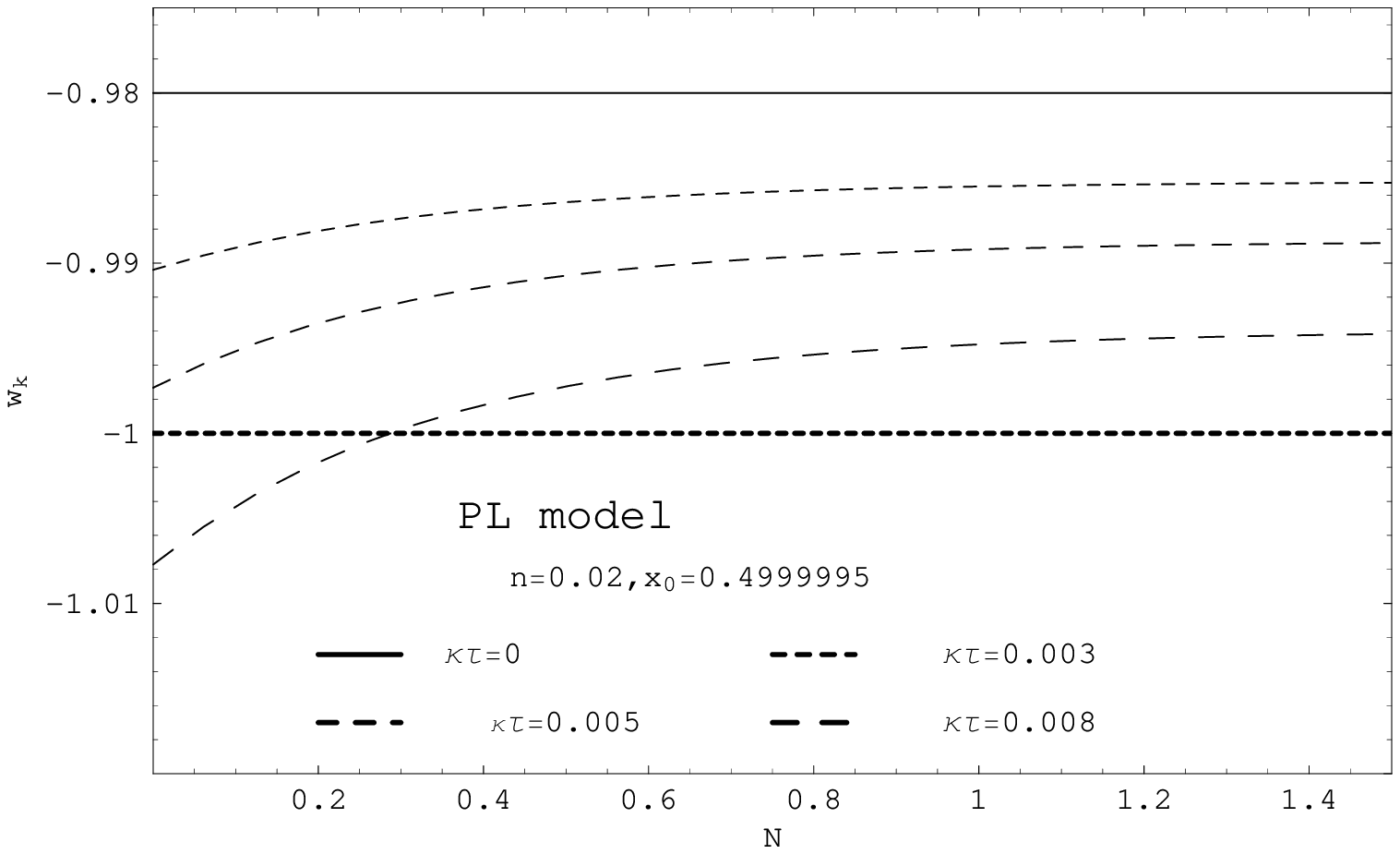}
 \hspace{0.1in}
 \includegraphics[height=0.9in,width=1.4in]{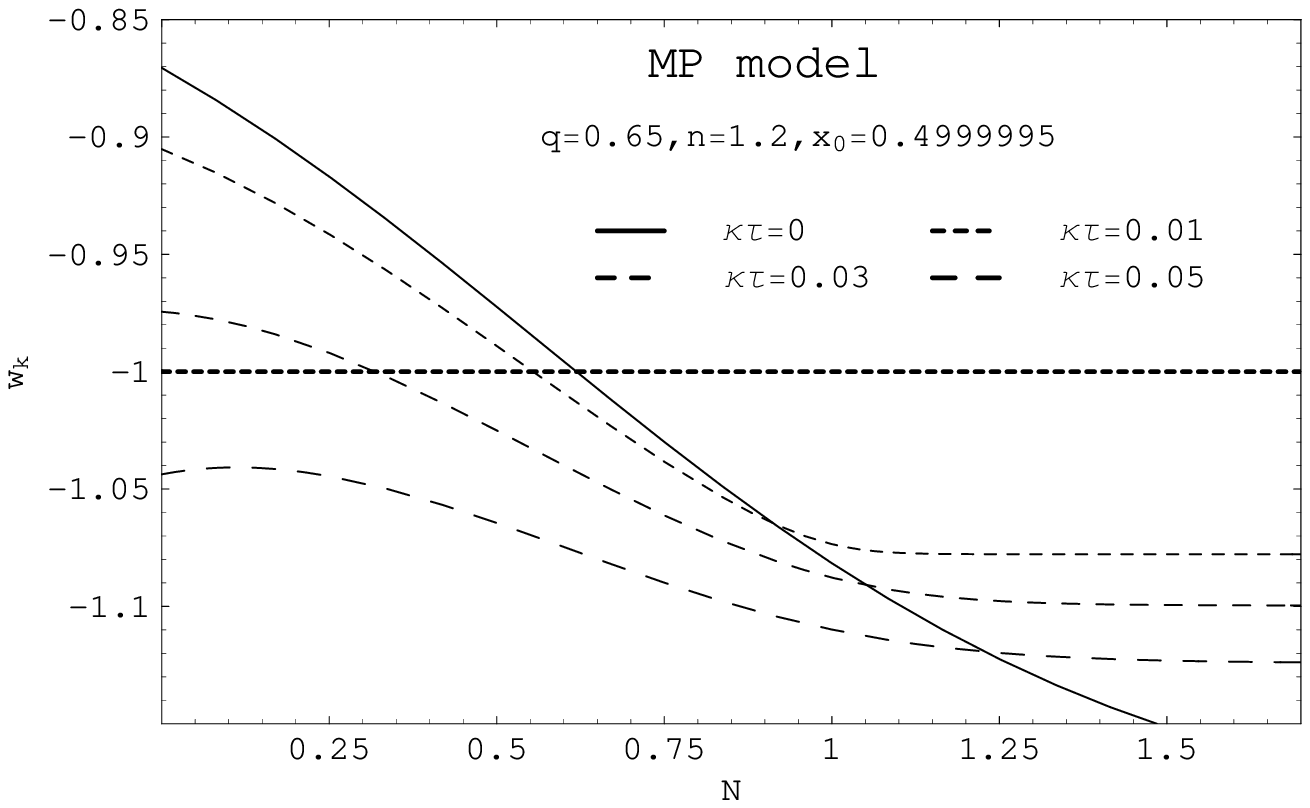}
 \hspace{0.1in}
 \includegraphics[height=0.9in,width=1.4in]{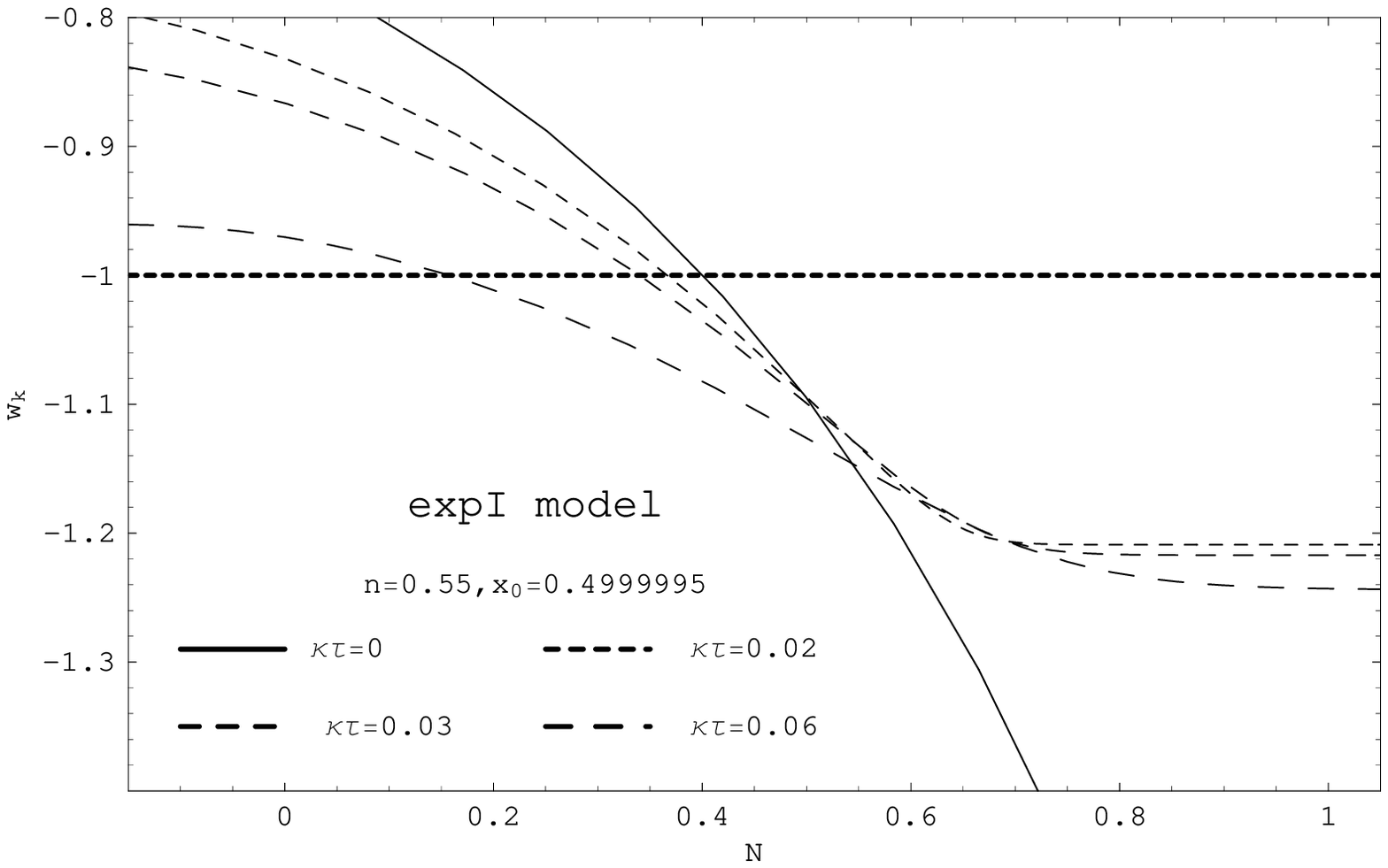}
\caption{In the $\zeta=\sqrt{3}\kappa^{-1}\tau H$ case, the
evolution of the equation-of-state parameters $w_{k}$ for different
values of $\kappa\tau$.\label{f1}}
\end{figure}

For the exp model, we have
\begin{eqnarray}\label{expI fOmega}
&&f(\Omega_{m})=\left[\frac{\ln\Omega_{m}}{(1-\sqrt{3}\kappa\tau+n\ln\Omega_{m})^{\sqrt{3}\kappa\tau}}\right]^{\frac{1}{3n(1-\sqrt{3}\kappa\tau)}}
\end{eqnarray}
and
\begin{eqnarray}\label{expI eq of state}
&&w_{k}=\frac{n\ln\Omega_{m}-\sqrt{3}\kappa\tau}{1-\Omega_{m}}.
\end{eqnarray}
We plot the evolution of $w_{k}$ for different values of
$\kappa\tau$ in Fig.~\ref{f1}. As can be seen, in viscous PL model,
$w_{k}$ is no longer a constant and it is dependent on time that can
cross the cosmological constant divide $w_{\Lambda}=-1$. For MP and
exp models, $w_{k}$ evolves more near $-1$ than the case without
viscosity.

 In the $\tau=0$ case, Eqs.(\ref{PL fOmega}) -(\ref{expI
eq of state}) reduce to the results of ordinary Cardassian model
without viscosity and Eq.(\ref{a expression}) is solvable for the
three models:
\begin{eqnarray}\label{exact solutions}
&&\Omega_{m}=\left\{\begin{array}{ll}
\frac{\frac{\Omega_{m,0}}{1-\Omega_{m,0}}(1+z)^{3(1-n)}}{1+\frac{\Omega_{m,0}}{1-\Omega_{m,0}}(1+z)^{3(1-n)}},
 & \textrm{for PL}\\
 \left[\frac{\frac{\Omega_{m,0}^{q}}{1-\Omega_{m,0}^{q}}(1+z)^{3qn}}{1+\frac{\Omega_{m,0}^{q}}{1-\Omega_{m,0}^{q}}(1+z)^{3qn}}\right]^{\frac{1}{q}},&\textrm{for MP}\\
 \\ \Omega_{m,0}^{(1+z)^{-3n}}.& \textrm{for
 exp}
 \end{array}\right.
\end{eqnarray}

\section{Autonomous system}
A general study of the phase space system of quintessence and
 phantom in FRW universe has been given in Ref.\refcite{phase space system}. For the viscous
 Cardassian cosmological dynamical system, the corresponding
 equations of motion can be written as
\begin{eqnarray}\label{H evolution}
&&\dot{H}=-\frac{\kappa^{2}}{2}\left(\rho_{m}\frac{dg(\rho_{m})}{d\rho_{m}}-3H\zeta\right),
\end{eqnarray}
\begin{eqnarray}\label{rhom evolution}
&&\dot{\rho_{m}}+3H\left(\rho_{m}-3H\zeta\frac{d\rho_{m}}{dg(\rho_{m})}\right)=0
\end{eqnarray}
and Eq.(\ref{Card FRW eq}). To analyze the dynamical system, we
rewrite the equations with the following dimensionless variables:
$x=\frac{\kappa^{2}\rho_{m}}{3H^{2}}$,$y=\frac{\kappa^{2}\zeta}{3H}$,$z=\frac{dg(\rho_{m})}{d\rho_{m}}$
and $N=\ln a$. The dynamical system can be reduced to
\begin{eqnarray}\label{Autonomous system}
&&\frac{dx}{dN}=\frac{9y}{z}-3x+3x^{2}z-9xy,\\
&&\frac{dy}{dN}=u+\frac{3}{2}xyz-\frac{9}{2}y^{2},
\end{eqnarray}
where $z$ and $u$ are both functions of $x$ and $y$. $z(x,y)$ is
determined by the specified form of $g(\rho_{m})$ and $u\equiv
\frac{\kappa^{2}}{3H^{2}}\frac{\partial\zeta}{\partial N}$ is
determined by the specified form of viscosity. The equation of state
can be expressed in terms of the new variables as
\begin{eqnarray}\label{w in new variables}
&&w_{k}=\frac{xz-3y-1}{1-x}
\end{eqnarray}
and the sound speed is
\begin{eqnarray}\label{sound speed}
&&c_{s}^{2}=\left\{\begin{array}{ll}
\frac{n(n-1)(1-x)}{n+(1-n)x},&\textrm{for PL}\\
\frac{n(1-x^{q})\left[n-1+(q-1)nx^{q}\right]}{nx^{q}+1-n}, & \textrm{for MP}\\
 \frac{n(1-n)\ln x+n^{2}(\ln x)^{2}}{1+n\ln x},& \textrm{for
 exp.}
 \end{array}\right.
\end{eqnarray}

\begin{figure}[!htbp]
\centering
\includegraphics[height=0.9in,width=1.4in]{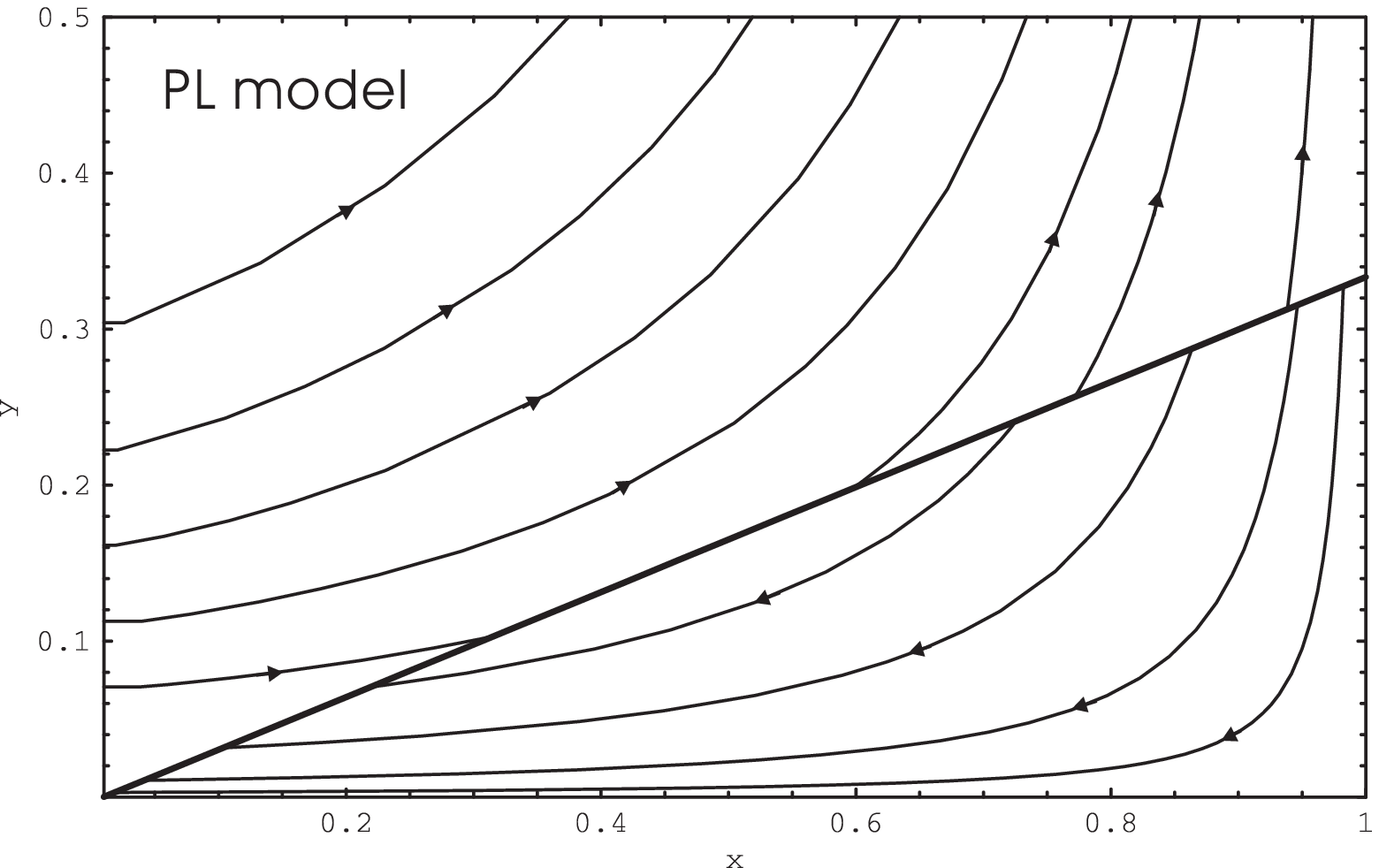}
 \hspace{0.1in}
 \includegraphics[height=0.9in,width=1.4in]{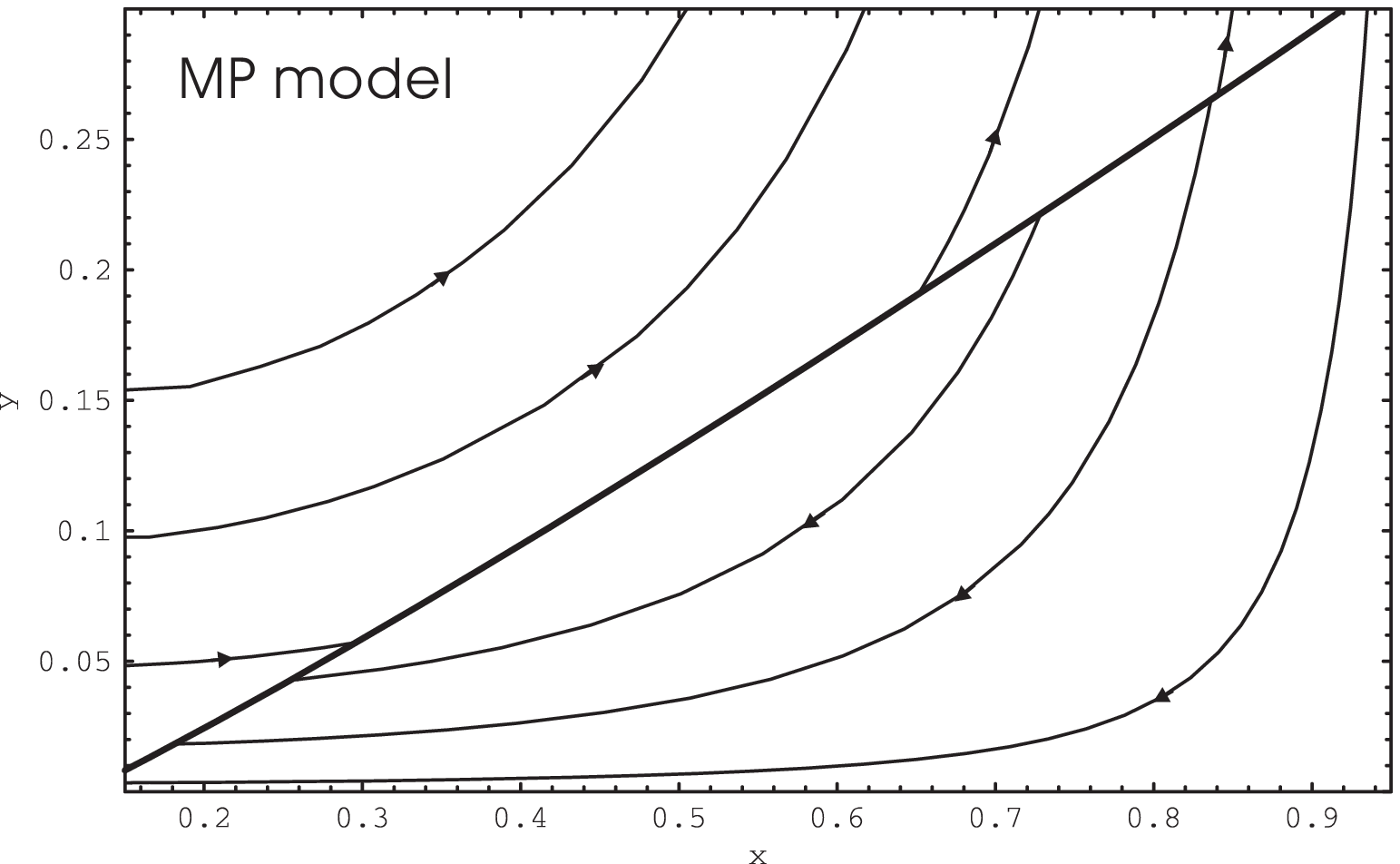}
 \hspace{0.1in}
 \includegraphics[height=0.9in,width=1.4in]{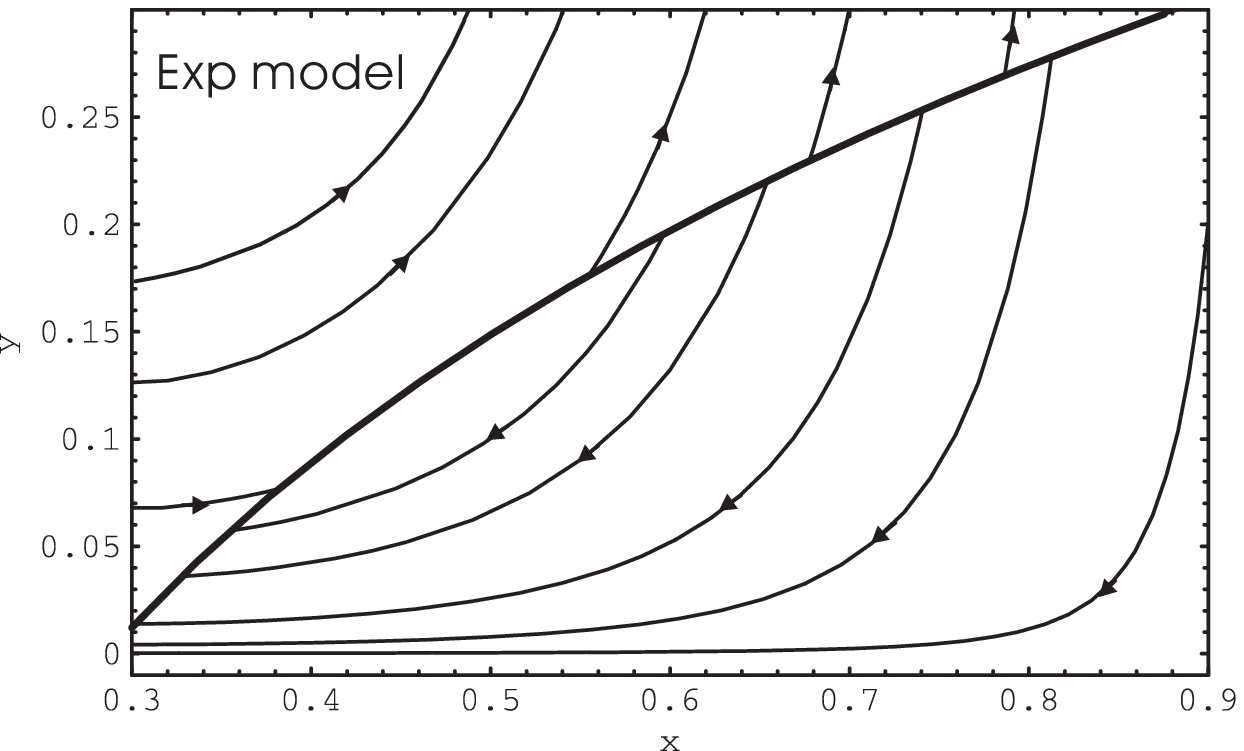}
\caption{In the $\zeta=\tau
\left(g(\rho)\right)^{\alpha+\frac{3}{2}}$ case, the phase diagrams
of $(x,y)$ for PL, MP and exp models. There is a singular curve
(bold line) which formed of critical points in the phase diagrams.
We choose the parameters to be $n=-0.01,\alpha=0.2$ for PL,
$q=1.15,n=1.1,\alpha=1.5$ for MP and $n=0.8,\alpha=2.5$ for exp,
respectively.\label{f2}}
\end{figure}

We choose $\zeta=\tau \left(g(\rho)\right)^{\alpha+\frac{3}{2}}$ and
show the critical points of the autonomous systems and their
properties in Table~\ref{ta1}. In Fig.~\ref{f2}, the phase diagrams
of the three models are given, respectively. It is worth noting that
there exists a singular curve (bold line) in the viscous
Cardassian's phase diagrams.

 \begin{center}
\begin{table}[!hbp]
\tbl{The critical points and their properties in the
$\zeta=\tau(g(\rho_{m}))^{\alpha+\frac{3}{2}}$ case.}
{\begin{tabular}{ ccc }
  \hline
  \hline
  \textbf{\scriptsize Critical points}$\scriptstyle(x,y)$& \textbf{\scriptsize Eigenvalues} & \textbf{\scriptsize Stability} \\
  \hline
  \hline
   &\qquad\qquad\qquad\qquad \textbf{\scriptsize PL model}&\\
    \hline
$\scriptstyle(0,0)$&$\left(\scriptstyle
3(n-1),\scriptstyle-3n(1+\alpha)\right)$&$\left\{
\begin{array}{ll}
\scriptstyle0<n<\frac{2}{3},\alpha>-1\\
\textrm{\scriptsize or}\scriptstyle\quad n<0,\alpha<-1
\end{array} \right.$\scriptsize ,stable \\
$\scriptstyle(1,0)$&$\left(\scriptstyle3(1-n),-3(1+\alpha)\right)$&\scriptsize unstable\\
$\scriptstyle\left(x_{0},\scriptstyle\frac{(1-n)x_{0}+n}{3}\right)$&$\left(\scriptstyle0,\scriptscriptstyle
3\left[(1-n)(2+\alpha)x_{0}+(1+\alpha)n-1+\frac{n}{n+(1-n)x_{0}}\right]\right)$&
$\left\{ \begin{array}{ll}
\scriptscriptstyle\frac{x_{0}}{x_{0}-1}<n<\frac{2}{3},\alpha<\frac{[1-4n+3n^{2}-2(1-n)^{2}x_{0}]x_{0}-n^{2}}{[n+(1-n)x_{0}]^{2}}\\
\textrm{\scriptsize or}\scriptscriptstyle
n<\frac{x_{0}}{x_{0}-1},\alpha>\frac{[1-4n+3n^{2}-2(1-n)^{2}x_{0}]x_{0}-n^{2}}{[n+(1-n)x_{0}]^{2}}
\end{array} \right.$\scriptsize stable \\
  \hline
  \hline
&\qquad\qquad\qquad\qquad \textbf{\scriptsize MP model}&\\
 \hline
$\scriptstyle(1,0)$&$\scriptstyle(-3(1+\alpha),3qn)$&\scriptsize unstable\\
$\scriptstyle\left(\left(\frac{n-1}{n}\right)^{\frac{1}{q}},0\right)$&$\scriptstyle(0,3q(n-1)-3)$&$\left\{
\begin{array}{ll}
\scriptstyle q(n-1)<1, &\textrm{\scriptsize stable}\\
\scriptstyle q(n-1)>1, &\textrm{\scriptsize unstable}
\end{array} \right.$\\
$\scriptstyle(x_{0},\scriptstyle\frac{1-n+nx_{0}^{q}}{3})$&$\left(\scriptstyle0,\scriptscriptstyle3\left[(1+\alpha)(1-n)+(1+q+\alpha)nx_{0}^{q}-\frac{qnx_{0}^{q}}{1-n+nx_{0}^{q}}\right]\right)$&$\left\{
\begin{array}{ll}
\scriptscriptstyle(1+\alpha)(1-n)+(1+n+\alpha)nx_{0}^{q}<\frac{nqx_{0}^{q}}{1-n+nx_{0}^{q}} \\
 \textrm{\scriptsize and}\scriptscriptstyle\quad 1-n+nx_{0}^{q}\neq0
\end{array} \right.$,\scriptsize stable\\
 \hline
  \hline
&\qquad\qquad\qquad\qquad \textbf{\scriptsize exp model}&\\
\hline
 $\scriptstyle(1,0)$&$\scriptstyle(3n,-3(1+\alpha))$&\scriptsize unstable\\
 $\scriptstyle(e^{-\frac{1}{n}},0)$&$\scriptstyle(0,3(n-1))$&$\left\{
\begin{array}{ll}
\scriptstyle0<n<1, &\textrm{\scriptsize stable}\\
\scriptstyle n>1, &\textrm{\scriptsize unstable}
\end{array} \right.$\\
$\scriptstyle\left(x_{0},\scriptstyle\frac{1+n\ln
x_{0}}{3}\right)$&$\left(\scriptstyle0,\scriptstyle3\left[1+\alpha+n+n(1+\alpha)\ln
x_{0}-\frac{n}{1+n\ln x_{0}}\right]\right)$&$\left\{
\begin{array}{ll}
\scriptstyle1+\alpha+n+n(1+\alpha)\ln x_{0}<\frac{n}{1+n\ln x_{0}} \\
 \textrm{\scriptsize and}\scriptstyle\quad 1+n\ln x_{0} \neq 0
\end{array} \right.$,\scriptsize stable\\
  \hline
\end{tabular}\label{ta1}}
\end{table}
\end{center}

\section{ Virialization}

The spherical collapse formalism developed by Gunn and
Gott\cite{Gunn} is a simple but powerful tool for studying the
growth of inhomogeneities and bound systems in the universe. It
describes how an initial inhomogeneity decouples from the expansion
of the universe and then expands slower, eventually reaches the
state of turnaround and collapses. Physically we assumes that the
collapse system goes through a virialization process and stabilizes
at a finite size. An important parameter of the spherical collapse
model is the ratio between the virialized radius and the turnaround
radius $\frac{R_{vir}}{R_{ta}}$\cite{Maor}.

Using energy conservation between virialization and
 turnaround, one can get
\begin{eqnarray}\label{energy conservation}
&&\left[U+T\right]_{vir}=U_{ta}
\end{eqnarray}
where $U$ is the potential energy, $T=\frac{R}{2}\frac{\partial
U}{\partial R}$ is the kinetic energy of the system. If the collapse
object is made up of only one energy component, the potential and
the potential energy within it are
\begin{eqnarray}\label{single potential}
&&\Phi(r)=-2\pi G(1+3w)\rho \left(R^{2}-\frac{r^{2}}{3}\right)
\end{eqnarray}
and
\begin{eqnarray}\label{single potential energy}
&&U=\frac{1}{2}\int \rho \Phi dV
\end{eqnarray}
where $R$ is the radius of the spherical collapse system. In the
Einstein-de Sitter universe, for a spherical perturbation with
conserved mass $M$, $\Phi(r)=\frac{GM}{2R^{3}}(r^{2}-3R^{2})$,
$U=-\frac{3GM^{2}}{5R}$, $T_{vir}=-\frac{U}{2}$, and the ratio of
virialization to turnaround is $\frac{R_{vir}}{R_{ta}}=\frac{1}{2}$.

\begin{figure}[!htbp]
\centering
\includegraphics[height=1.4in,width=2.2in]{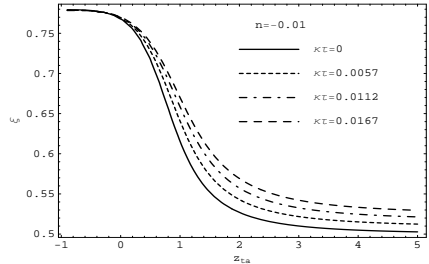}
\caption{In the viscous PL model with $\zeta=\sqrt{3}\kappa^{-1}\tau
H$, $\xi$ as a function of $z_{ta}$ for different bulk viscosity
coefficients $\tau$.\label{f3}}
\end{figure}

In the PL model without bulk viscosity, we have
\begin{eqnarray}\label{pow law model total potential}
&&\Phi=-2G\pi
\rho_{m}\left(R^{2}-\frac{r^{2}}{3}\right)\left[1+\left(\frac{\rho_{m}}{\rho_{card}}\right)^{n-1}\right]\left\{1+3\left[n-1+\frac{(1-n)\rho_{m}}{\rho_{m}+\rho_{card}\left(\frac{\rho_{m}}{\rho_{card}}\right)^{n}}\right]\right\}\nonumber\\
\end{eqnarray}
and
\begin{eqnarray}\label{pow law model total potential energy}
&&U=-\frac{16}{15}G\pi^{2}R^{5}\left[\rho_{m}+\rho_{card}\left(\frac{\rho_{m}}{\rho_{card}}\right)^{n}\right]\left[\rho_{m}+(3n-2)\rho_{card}\left(\frac{\rho_{m}}{\rho_{card}}\right)^{n}\right]\;.\nonumber\\
\end{eqnarray}
Using the relation $\rho_{m,vir}=\rho_{m,ta}\xi^{-3}$ where
$\xi\equiv \frac{R_{vir}}{R_{ta}}$, Eq.(\ref{energy conservation})
can be written as
\begin{eqnarray}\label{pow law model full system virial relation of x and z}
&&\left\{2\xi\left[1+(1+z_{ta})^{3n-3}(\Omega_{m,0}^{-1}-1)\right]\left[(3n-2)(1+z_{ta})^{3n-3}(\Omega_{m,0}^{-1}-1)+1\right]-1\right\}\xi^{6n}+\nonumber\\
&&(9n^{2}-15n+4)(1+z_{ta})^{3n-3}(\Omega_{m,0}^{-1}-1)\xi^{3+3n}+(3n-2)(6n-7)(1+z_{ta})^{6n-6}(\Omega_{m,0}^{-1}-1)^{2}\xi^{6}\nonumber\\
&&=0\;.
\end{eqnarray}
It is difficult to find analytical solution for Eq.(\ref{pow law
model full system virial relation of x and z}), therefore we have to
investigate it numerically. The result is shown in Fig.~\ref{f3}
(solid line). In the viscous PL model with
$\zeta=\sqrt{3}\kappa^{-1}\tau H$, same problem becomes more
complex. As the function of scale factor $a$, $\rho_{m}$ is
determined by
\begin{eqnarray}\label{relation of x and rho for PL}
&&\left[\sqrt{3}\kappa\tau\left(1+\left(\frac{\rho_{card}}{\rho_{m,0}}\right)^{1-n}\right)-1-n\left(\frac{\rho_{card}}{\rho_{m,0}}\right)^{1-n}\right]\left(\frac{a}{a_{0}}\right)^{\frac{3(\sqrt{3}\kappa\tau-1)(\sqrt{3}\kappa\tau-n)}{\sqrt{3}\kappa\tau}}\nonumber\\
&&=(\sqrt{3}\kappa\tau-1)\left(\frac{\rho_{m}}{\rho_{m,0}}\right)^{1-\frac{n}{\sqrt{3}\kappa\tau}}+(\sqrt{3}\kappa\tau-n)\left(\frac{\rho_{card}}{\rho_{m,0}}\right)^{1-n}\left(\frac{\rho_{m}}{\rho_{m,0}}\right)^{n-\frac{n}{\sqrt{3}\kappa\tau}}.
\end{eqnarray}
Combing (\ref{relation of x and rho for PL}) and (\ref{energy
conservation}), we obtain the relation between $\xi$ and turnaround
redshift $z_{ta}$. In Fig.~\ref{f3}, we show $\xi$ as a function of
$z_{ta}$ for different bulk viscosity coefficients. As can be seen,
in Cardassian models, the ratio $\frac{R_{vir}}{R_{ta}}$ is always
larger than $\frac{1}{2}$, and it get larger and larger with the
evolution of the universe. It means that collapse process will be
harder and harder to occur ($\xi\rightarrow1$). Furthermore, $\xi$
becomes larger when $\tau$ takes larger value.

\section{Fit the model parameters to supernovae data}
In general, the approach towards determining the expansion history
$H(z)$ is to assume an arbitrary ansatz for $H(z)$ which is not
necessarily physically motivated  but is specially designed to give
a good fit to the data for $d_L (z)$. Given a particular
cosmological model for $H(z; a_1, ... ,a_n)$ where $a_1, ...,a_n$
are model parameters, the maximum likelihood technique can be used
to determine the best fit values of parameters as well as the
goodness of the fit of the model to the data. The technique can be
summarized as follows: The observational data consist of $N$
apparent magnitudes $m_i (z_i)$ and redshifts $z_i$ with their
corresponding errors $\sigma_{m_i}$ and $\sigma_{z_i}$. These errors
are assumed to be gaussian and uncorrelated. Each apparent magnitude
$m_i$ is related to the corresponding luminosity distance $d_L$ by
\begin{eqnarray}\label{mz}
&&m(z)=M+5\log_{10}\left[\frac{d_{L}(z)}{Mpc}\right]+25,
\end{eqnarray}
where $M$ is the absolute magnitude. For the distant SNeIa, one can
directly observe their apparent magnitude $m$ and redshift $z$,
because the absolute magnitude $M$ of them is assumed to be
constant, i.e., the supernovae are standard candles. Obviously, the
luminosity distance $d_L (z)$ is the `meeting point' between the
observed apparent magnitude $m(z)$ and the theoretical prediction
$H(z)$. Usually, one define distance modulus $\mu(z)\equiv m(z)-M$
and express it in terms of the dimensionless `Hubble-constant free'
luminosity distance $D_L$ defined by$D_L (z) = {{H_0 d_L (z)}/ c}$
as
\begin{eqnarray}\label{muz}
&&\mu(z)=5\log_{10}\left(D_{L}(z)\right)+\mu_{0},
\end{eqnarray}
where the zero offset $\mu_0$ depends on $H_0$ (or $h$) as
\begin{eqnarray}\label{mu0}
&&\mu_{0}=5\log_{10}\left(\frac{cH_{0}^{-1}}{Mpc}\right)+25=-5\log_{10}h+42.38.
\end{eqnarray}
The theoretically predicted value $D_L^{th} (z)$ in the context of a
given model $H(z;a_1,...,a_n)$ can be described by
\begin{eqnarray}\label{DLth}
&& D_L^{th} (z) = (1+z) \int_0^z dz' \; {{H_0}\over
{H(z';a_1,...a_n)}}.
\end{eqnarray}
If we assume prior to the parameters $\Omega_{m,0}$ , the viscous
Cardassian models predict a specific form of the Hubble parameter
$H(z)$ as a function of redshift $z$ in terms of two parameters $n$
 and $\tau$ for the viscous PL model and exp model. Therefore, the best fit values for the parameters $(n,
 \tau)$ are found by minimizing the quantity
\begin{eqnarray}\label{chi2}
&&\chi^2 (n,\tau)=\sum_{i=1}^N
 \frac{\left[\mu^{{obs}}(z_i)-5\log_{10}D_L^{{th}}(z_i;n,\tau)-\mu_0\right]^2}{\sigma_i^2}.
\end{eqnarray}
Since the nuisance parameter $\mu_0$ is model-independent, its value
from a specific good fit can be used as consistency test of the data
\cite{SNeIa} and one can choose a priori value of it (equivalently,
the value of dimensionless Hubble parameter $h$) or marginalize over
it thus obtaining
\begin{eqnarray}\label{chi22}
&&\widetilde{\chi}^2(n,\tau)=A(n,\tau)
-\frac{B(n,\tau)^2}{C}+\ln\left(\frac{C}{2\pi}\right),
\end{eqnarray}
where
\begin{eqnarray}\label{A}
&& A(n,\tau)=\sum_{i=1}^N
 \frac{\left[\mu^{{obs}}(z_i)-5\log_{10}D_L^{{th}}(z_i;n,\tau)\right]^2}{\sigma_i^2},
\end{eqnarray}
\begin{eqnarray}\label{B}
&& B(n,\tau)=\sum_{i=1}^N
 \frac{\left[\mu^{{obs}}(z_i)-5\log_{10}D_L^{{th}}(z_i;n,\tau)\right]}{\sigma_i^2},
\end{eqnarray}
and
\begin{eqnarray}\label{C}
   C=\sum_{i=1}^N
 \frac{1}{\sigma_i^2}.
\end{eqnarray}
In the latter approach, instead of minimizing ${\chi}^2(n,\tau)$,
one can minimize $\widetilde{\chi}^2(n,\tau)$ which is independent
of $\mu_0$.

\begin{figure}[!htbp]
\centering
\includegraphics[height=1.4in,width=2.2in]{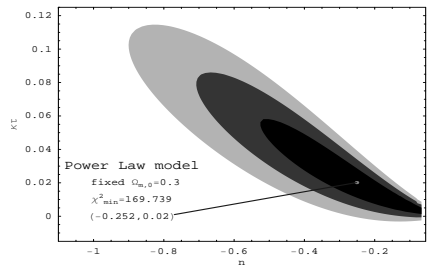}
\caption{The $68.3\%$,$95.4\%$ and $99.7\%$ confidence contours of
parameters $n$ and $\kappa\tau$ using the Gold SNeIa dataset and
marginalizing over the model-independent parameter $\mu_{0}$ with a
prior $\Omega_{m,0}=0.3$ in PL model.\label{f4}}
\end{figure}

We now apply the above described maximum likelihood method for the
viscous PL model using Gold dataset which is one of the reliable
publish dataset consisting of 206 SNeIa (N=206)\cite{SNeIa}. In
Fig.~\ref{f4}, contours with $68.3\%$,$95.4\%$ and $99.7\%$
confidence level are plotted, in which we take a marginalization
over the model-independent parameter $\mu_{0}$. The best fit as
showed in the figure corresponds to $n=-0.252$ and
$\kappa\tau=0.02$, and the minimum value of $\chi^{2}$ is $169.739$.

\begin{figure}[!htbp]
\centering
\includegraphics[height=1.4in,width=2.2in]{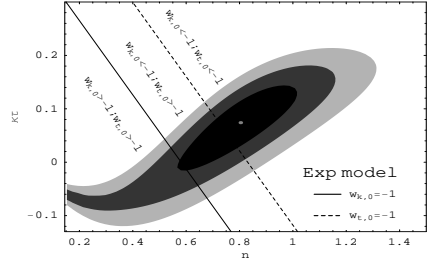}
\caption{The $68.3\%$,$95.4\%$ and $99.7\%$ confidence contours of
parameters $n$ and $\kappa\tau$ using the Gold SNeIa dataset and
marginalizing over the model-independent parameter $\mu_{0}$ with a
prior $\Omega_{m,0}=0.3$ in exp model. $w_{k,0}$ and $w_{t,0}$ are
equation-of-state parameters of $\rho_{k}$ and $g(\rho_{m})$ at
present time, respectively.\label{f5}}
\end{figure}

In Fig.~\ref{f5}, contours with  $68.3\%$,$95.4\%$ and $99.7\%$
confidence level are plotted for the viscous exp model. The best fit
is
 $n=0.8$ and $\kappa\tau=0.074$, and $\chi^{2}_{min}=169.857$. Obviously,
 the allowed ranged of the parameters $n$ and $\tau$ favor that there exists an
effective phantom energy in the universe.

As for the viscous MP model, using the above marginalization method,
we find the minimum value of $\chi^{2}(q,n,\tau)$ is 169.823 and the
corresponding best fit value of parameters are $q=0.233$, $n=3.853$,
and $\kappa\tau=0.06$.

\section{Discussion and conclusion}
In above sections, we have investigated the viscous Cardassian
cosmology for the expanding universe, assuming that there is a bulk
viscosity in the cosmic fluid. We consider two solvable cases: (i)
$\zeta=\sqrt{3}\kappa^{-1}\tau H$; (ii) $\zeta=\tau
\left(g(\rho)\right)^{\alpha+\frac{3}{2}}$. In case (i), there exist
exact solutions of $a$ as a function of $\Omega_{m,0}$ for PL, MP
and exp models. Contrary to the naive PL model, in viscous PL model,
the effective equation-of-state parameter $w_{k}$ is no longer a
constant and it is dependent on time that can cross the cosmological
constant divide $w_{\Lambda}=-1$ from $w_{k}>-1$ to $w_{k}<-1$.
Other models possess with similar characteristics. For MP and exp
models, $w_{k}$ evolves more near $-1$ than the case without
viscosity. Moreover, the dynamical analysis indicates that there
exists a singular curve in the phase diagram of viscous Cardassian .

 On the other hand, the bulk viscosity can effect the collapse
process of a bound system in the universe. We consider the
virialization of a spherical collapse model using the spherical
collapse formalism in PL Cardassian. The numerical result indicates
that the parameter $\xi=\frac{R_{vir}}{R_{ta}}$ is always larger
than $\frac{1}{2}$ and get larger and larger with the evolution of
the universe. Furthermore, $\xi$ becomes larger when $\tau$ is
increasing that means the bulk viscosity retards the progress of
collapse system. Obviously, the bulk viscosity should be small to
insure that the viscous cosmology theory isn't in contradiction with
the galaxy formation theory.

Using maximum likelihood technique, we constrain the parameters of
viscous Cardassian models from the supernova data. If we assume
prior that $\Omega_{m,0}=0.3$ as indicated by the observation about
mass function of galaxies, the best fits of $\kappa\tau$ are $0.02$,
$0.06$ and $0.07$ for PL, MP and exp models, respectively. In the
following works, we plan to use other cosmological and astrophysical
observations such as CMB, BAO and LSS to further constrain the
viscous Cardassian parameters and the bulk viscosity coefficient.

\section*{Acknowledgments}

This work is supported by National Natural Science Foundation of
China.

\end{document}